\documentclass[12pt]{article}
\usepackage{geometry}
\usepackage[sort&compress,square,comma]{natbib}
\usepackage{amsmath}
\usepackage{amssymb}
\usepackage{amsmath}
\usepackage{booktabs}
\usepackage{amsthm}
\usepackage{epsfig}
\usepackage{psfrag}

\numberwithin{equation}{section}

\def\appendix#1{
  \addtocounter{section}{1}
  \setcounter{equation}{0}
  \renewcommand{\thesection}{\Alph{section}}
 \section*{Appendix \thesection\protect\indent \parbox[t]{11.715cm} {#1}}
  \addcontentsline{toc}{section}{Appendix \thesection\ \ \ #1}
  }

\renewcommand{\thefootnote}{\fnsymbol{footnote}}

\newcommand{\CP}{{\mathbb C \mathbb P}^3}
\newcommand{\AD}{AdS_5\times S^5}
\newcommand{\Ad}{AdS_4\times {\mathbb C \mathbb P}^3}
\newcommand{\be}{\begin{equation}}
\newcommand{\ee}{\end{equation}}
\newcommand{\ba}{\begin{aligned}}
\newcommand{\ea}{\end{aligned}}

\def\m1{\left(-1\right)^{F_i}}

\makeatletter
\def\sla@#1#2#3#4#5{{%
  \setbox\z@\hbox{$\m@th#4#5$}%
  \setbox\tw@\hbox{$\m@th#4#1$}%
  \dimen4\wd\ifdim\wd\z@<\wd\tw@\tw@\else\z@\fi
  \dimen@\ht\tw@
  \advance\dimen@-\dp\tw@
  \advance\dimen@-\ht\z@
  \advance\dimen@\dp\z@
  \divide\dimen@\tw@
  \advance\dimen@-#3\ht\tw@
  \advance\dimen@-#3\dp\tw@
  \dimen@ii#2\wd\z@  \raise-\dimen@\hbox to\dimen4{%
    \hss\kern\dimen@ii\box\tw@\kern-\dimen@ii\hss}%
  \llap{\hbox to\dimen4{\hss\box\z@\hss}}}}
\def\slashed#1{%
  \expandafter\ifx\csname sla@\string#1\endcsname\relax
    {\mathpalette{\sla@/00}{#1}}%
  \else
    \csname sla@\string#1\endcsname
  \fi}
\makeatother

\newcommand{\E}{{\cal E}}

\newcommand{\beq}{\begin{equation}}
\newcommand{\eeq}{\end{equation}}
\newcommand\beqa{\begin{eqnarray}}
\newcommand\eeqa{\end{eqnarray}}
\newcommand\bea{\begin{array}}
\newcommand\eea{\end{array}}
\newcommand{\COMMENT}[1]{{}}
\newcommand{\nn}{\nonumber}
\newcommand{\neqa}{\nonumber\end{eqnarray}}
\newcommand{\la}{\label}

\newcommand{\color}[1]{}

\newcommand{\eq}[1]{Eq.(\ref{#1})}
\newcommand{\eqs}[2]{Eqs.(\ref{#1},\ref{#2})}

\def\({\left(}
\def\){\right)}
\def\[{\left[}
\def\]{\right]}

\def\<{\langle}
\def\>{\rangle}

\newcommand{\beqn}{\begin{eqnarray}}
\newcommand{\eeqn}{\end{eqnarray}}
\newcommand{\fr}{\frac}
\newcommand{\nnr}{\nonumber\\}

\begin{document}


\thispagestyle{empty}
\begin{flushright}\footnotesize
\texttt{LPTENS 08/NN}\\
\texttt{SPhT-t08/NNN}\\

\texttt{ITEP-TH-35/08}\\

\vspace{2.1cm}
\end{flushright}

\renewcommand{\thefootnote}{\fnsymbol{footnote}}
\setcounter{footnote}{0}
\setcounter{figure}{0}
\begin{center}
{\Large\textbf{\mathversion{bold} Comment on the Scaling Function in $\Ad$}\par}

\vspace{2.1cm}

\textrm{Nikolay Gromov$^{\alpha}$, Victor Mikhaylov$^{\beta}$}
\vspace{1.2cm}

\textit{$^{\alpha}$ Service de Physique Th\'eorique,
CNRS-URA 2306 C.E.A.-Saclay, F-91191 Gif-sur-Yvette, France;
Laboratoire de Physique Th\'eorique de
l'Ecole Normale Sup\'erieure et l'Universit\'e Paris-VI,
Paris, 75231, France;
St.Petersburg INP, Gatchina, 188 300, St.Petersburg, Russia } \\
\texttt{nikgromov@gmail.com}
\vspace{3mm}

\textit{$^{\beta}$ Institute for Theoretical and Experimental Physics,\\
B.~Cheremushkinskaya 25, Moscow 117259, Russia; Moscow Institute of Physics and Technology,
Institutsky per. 9, 141 700, Dolgoprudny, Russia}\\
\texttt{victor.mikhaylov@gmail.com}
 \vspace{3mm}


\par\vspace{1cm}

\textbf{Abstract}\vspace{5mm}
\end{center}

\noindent

The folded spinning string in $AdS_3$ gives us an important insight into $AdS/CFT$ duality.
Recently its one-loop energy was analyzed in the context of $AdS_4/CFT_3$ by McLoughlin and Roiban
arXiv:0807.3965, by Alday, Arutyunov and Bykov arXiv:0807.4400 and by 
Krishnan arXiv:0807.4561.
They computed the spectrum of the fluctuations around the classical solution.

In this paper we reproduce their results using the algebraic curve technique
and show that under some natural resummation of the fluctuation energies the one-loop energy
agrees perfectly with the predictions of arXiv:0807.0777.
This provides a further support of the all-loop Bethe equations and of the $\Ad$ algebraic curve
developed in arXiv:0807.0437.

\vspace*{\fill}

\setcounter{page}{1}
\renewcommand{\thefootnote}{\arabic{footnote}}
\setcounter{footnote}{0}

\newpage



\section{Introduction}
Integrability in $AdS/CFT$ duality \cite{Maldacena:1997re} is an exciting subject in the modern theoretical physics.
The integrability of 4D Yang-Mills theory was first discussed in \cite{Lipatov:1993yb,Faddeev:1994zg}.
An intensive development started after the seminal paper by Minahan and Zarembo \cite{Minahan:2002ve} where the long single trace operators of ${\cal N}=4$ Super Yang-Mills theory, dual to the superstring in the $AdS_5\times S^5$
background, were mapped onto integrable spin chains. The string theory was shown to be
classically integrable \cite{Bena:2003wd} and is widely believed to be integrable at the quantum level as well. Many attempts were made towards complete understanding of the spectrum of this system. In particular,  the asymptotic Bethe ansatz proposed in \cite{Staudacher:2004tk,Beisert:2005fw,Beisert:2006ez} was an important step in this direction.

Recently, a new duality was discovered between ${\cal N}=6$ super Chern-Simons theory
with $U(N)\times U(N)$ gauge group at level $k$
and superstring theory in the $AdS_4\times {\mathbb C\mathbb P}^3$ background
in the large $N$ limit with the `t Hooft coupling $\lambda=N/k$ kept fixed \cite{Aharony:2008ug,Benna:2008zy}.
Amazingly, the gauge side of the duality also exhibits integrability \cite{Minahan:2008hf} (see also
\cite{Gaiotto:2008cg,Bak:2008cp}).
The string theory turns out to be classically integrable as well \cite{Arutyunov:2008if,Stefanski:2008ik}.
The all-loop Bethe equations were proposed recently in \cite{Gromov:2008qe} and have already passed several
independent tests \cite{Astolfi:2008ji,Ahn:2008aa}.

An important consequence of the classical integrability is an existence of the finite gap algebraic curve
\cite{Kazakov:2004qf,Beisert:2005bm}. For the
$AdS_4\times \CP$ superstring theory it was constructed in \cite{Gromov:2008bz}.
The algebraic curve describes all classical solutions in a unified
gauge-invariant way and can be also used for computation of the spectrum of quantum fluctuations
around a given classical solution \cite{Gromov:2007aq}. We will use this technique to compute
fluctuation frequencies around a particularly important folded string classical solution. 
These frequencies were obtained recently in \cite{McLoughlin:2008ms,Alday:2008ut,Krishnan:2008zs} by a direct diagonalization of the string action expanded
around the classical solution. The match of our results provides a nontrivial test of the algebraic curve
construction \cite{Gromov:2008bz}.

The folded string is rotating in $AdS_3$ with large spin $S$
and with angular momentum $J\sim \log S$ in ${\mathbb C\mathbb P}^3$. The difference between its energy and spin scales as $J$ \cite{Gubser:2002tv,Belitsky:2006en} for all values of the `t Hooft coupling $\lambda$ \cite{Aharony:2008ug}.
A similar solution in the context of $AdS_5/CFT_4$ duality has been already well studied for all
values of $\lambda$ \cite{Kotikov:2006ts,Alday:2007qf,Kostov:2007kx,Beccaria:2007tk,Kostov:2008ax,Casteill:2007ct,Belitsky:2007kf,Roiban:2007jf,
Alday:2007mf,Basso:2007wd,Roiban:2007dq,Roiban:2007ju,Fioravanti:2008rv,Freyhult:2007pz,Basso:2008tx,Fioravanti:2008ak,Buccheri:2008ap,Gromov:2008en,Beccaria:2008nf}. In \cite{Gromov:2008qe} an equation describing the $sl(2)$ sector of
the $\Ad$ string was proposed:
\beq
\(\frac{x_k^+}{x_k^-}\)^J=-\prod_{j\neq k}^S\(\frac{x_k^+-x_j^-}{x_k^--x_j^+}\)^{-1}\frac{1-1/(x_k^+x_j^-)}{1-1/(x_k^-x_j^+)}\sigma^2_{\rm BES}(u_k,u_j)\;.
\eeq
It describes, in particular, the folded string. One can see that the only difference with $\AD$ string Bethe ansatz \cite{Arutyunov:2004vx} is a minus sign on the r.h.s.
It produces a simple redefinition of a scaling parameter
\beq
\ell_{AdS_5}=\frac{J}{4g\log S}\;\;\to\;\;\ell_{AdS_4}=\frac{J}{2h\log S}\;,
\eeq
where $g=\sqrt{\lambda_{YM}}/4\pi$ and $h(\lambda_{CS})=\sqrt{\lambda_{CS}/2}+{\cal O}(1/\sqrt{\lambda_{CS}})$ for
large $\lambda_{CS}$ and $h(\lambda_{CS})=\lambda_{CS}+{\cal O}(\lambda^3_{CS})$ for small $\lambda_{CS}$
\cite{Gaiotto:2008cg,Grignani:2008is,
Nishioka:2008gz,Shenderovich:2008bs}.
This implies, in turn, that once the energy of the string in $\AD$ is given by
\beq
\gamma_{\rm YM}=J\frac{f(g,\ell_{AdS_5})}{\ell_{AdS_5}}\;,
\eeq
the energy of the folded string in $\Ad$ reads~\cite{Gromov:2008qe}
\beq\la{fym}
\gamma_{\rm CS}=J\frac{f(h,\ell_{AdS_4})}{\ell_{AdS_4}}\;,
\eeq
where the one-loop scaling function $f$ is given by~\cite{Frolov:2006qe}
\beq\la{f1loop}
f(h,\ell)\simeq\(\sqrt{\ell^2+1}-\ell\)
+\frac{\sqrt{\ell^2+1}-1+2(\ell^2+1)\log\(1+\frac{1}{\ell^2}\)-(\ell^2+2)\log\frac{\sqrt{\ell^2+2}}
{\sqrt{\ell^2+1}-1}}{4\pi h\sqrt{\ell^2+1}}\;.
\eeq
In the limit $\ell \to 1$ one finds
\beq
f(h,0)\simeq 1-\frac{3\log 2}{4\pi h}\;.
\eeq

Recently the folded string solution at one-loop was analyzed
in $\Ad$ \cite{McLoughlin:2008ms,Alday:2008ut,Krishnan:2008zs} starting from a superstring action.
In these papers the spectrum of fluctuations
around the folded string was computed and summed up into a one-loop shift.
A disagreement with \eqs{fym}{f1loop} was stated.
According to these works, in the limit $\ell\to 0$ the scaling function is equal to $-\frac{5\log 2}{4\pi h}$.

In this paper we  reproduce the results for the fluctuation energies from the finite gap algebraic curve and
propose a natural regularization of the sum of fluctuations
leading precisely to the conjectured result (\ref{fym},\ref{f1loop}).


\section{Algebraic Curve for the Folded String}
In this section we recall the method
of finding the fluctuation energies directly from the algebraic curve \cite{Gromov:2007aq}.
The algebraic curve allows us to compute these fluctuation energies disregarding a particular
parametrization and gauge fixing of the superstring action. This provides us with a universal framework
for computation of the one-loop corrections to the classical energies for superstrings living in different
integrable backgrounds. The finite-gap solutions with any number of cuts can be treated on equal grounds \cite{Gromov:2008}. Moreover, the calculations are much simpler then by the other methods. In particular, for the folded string solutions all the fluctuations
can be obtained just from comparing the algebraic curves for the $\AD$
and $\Ad$ superstrings.

We will start with describing the properties of the algebraic curves. The quasimomenta
for the classical solution of string sigma-model in $AdS_4\times\CP$ are not independent, namely,
\beq
\{q_1,~q_2,~q_3,~q_4,~q_5\}=-\{q_{10},~q_9,~q_8,~q_7,~q_6\}\,.\label{q-q}
\eeq
Here the quasimomenta $q_1,~q_2,~q_9,~q_{10}$ are responsible for the $AdS_4$ part of sigma-model
and the others for the $\CP$ part \cite{Gromov:2008bz}.

Since the motion of the folded string is constrained to $AdS_3\times S^1$ subspace, the
$AdS$-quasimomenta $q_1,~q_2$ coincide with those for the folded
string in $\AD$. They have two cuts shared with the functions $q_{10}$ and $q_9$, respectively.
The quasimomenta in $\CP$ are the same as in the point-like string limit (BMN limit), because the motion in this subspace is trivial,
\beqn
&&q_3(x)=q_4(x)=-q_8(x)=-q_7(x)=\fr{2\pi\nu x}{x^2-1}\,,\nnr
&&q_5(x)=-q_6(x)=0\,.\label{cpq-s}
\eeqn

\begin{figure}[t]
\epsfxsize=15cm
\centerline{\epsfbox{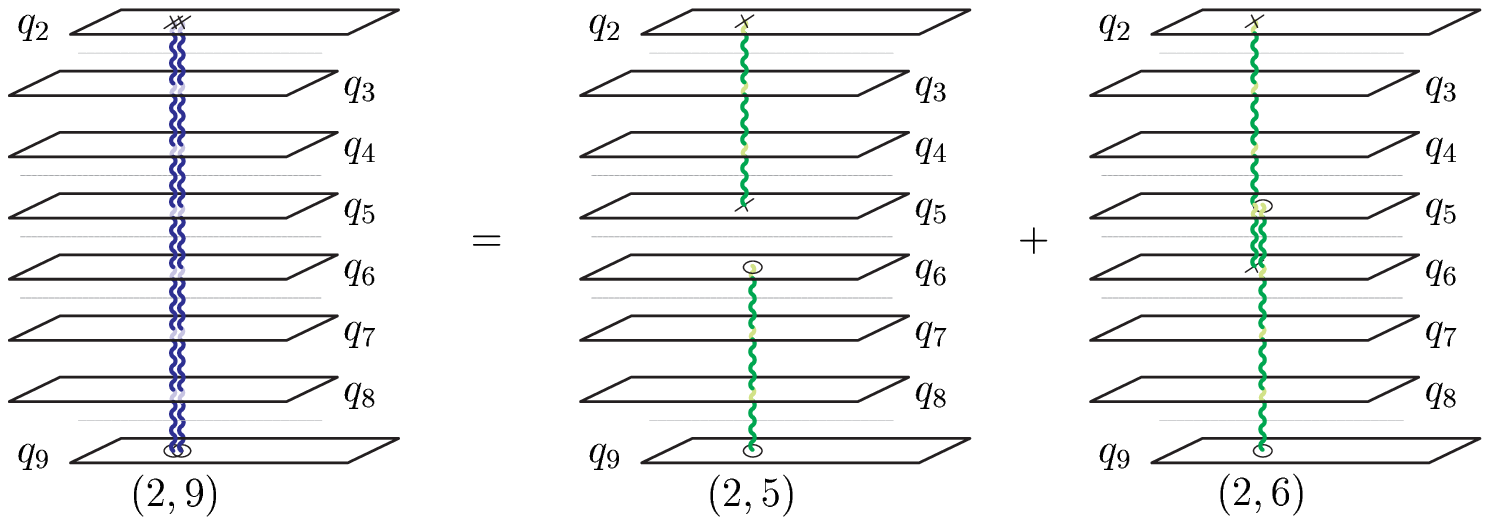}}
\caption{\label{pic29}\small Equation relating fermionic and bosonic fluctuations (2,5) and (2,9). Adding a pole with double residue between sheets 2 and 9 is equivalent to
adding a (2,5) pole plus a (2,6) pole. The fermionic fluctuations with polarizations (2,5) and (2,6) are equivalent since $q_5=q_6$.}
\end{figure}

Let us briefly recall the idea of quantization of the string using the algebraic curve.
In order to obtain quasiclassical frequencies one has to perturb the classical curve and find the corresponding shift of
the energy. These perturbations of the quasimomenta are implemented by adding some poles playing a role of infinitesimal cuts. We denote the polarization of the excitation by a pair of numbers (i,j) corresponding to the sheets between which the pole is added. If one sheet corresponds to
$AdS$ and another to $\mathbb C\mathbb P$, the fluctuation is fermionic, otherwise it is bosonic.

The residues of the poles are fixed by quasiclassical quantization condition and their positions are determined by the equations
\beq
q_i(x_n^{(i,j)})-q_j(x_n^{(i,j)})=2\pi n_{(i,j)}\,.\label{positions}
\eeq
Thus the calculation splits into two steps: (i) calculating the response of the energy $\delta E=\Omega_{ij}(x)$ to insertion of a pole at some point $x$ and (ii) solving the equations \ref{positions}. Then the fluctuation energy is simply $\delta E^{(i,j)}_n=\Omega_{ij}(x_n^{(i,j)})$.

In our case one can notice that all the functions $q_i(x)$ have already appeared in the $\AD$ algebraic curve, except for the
trivial functions $q_5$ and $q_6$. This allows us to write immediately most of the fluctuation frequencies. For example, the fermionic excitation energy $\delta E_n^{(2,7)}$ is obtained by adding a pole between a sheet with two cuts in the physical
region and a BMN-like sheet (i.e. containing only two poles at $\pm1$). In the $\AD$ such fluctuation has polarization $(\hat{2},\tilde{3})$
and hence its frequency is equal to $\omega_n^F$ given in Tab.\ref{tab1}. In the same way one finds the other $\delta E_n^{(i,j)}$'s, not involving
the sheets 5 and 6, which we call ``heavy" fluctuations.

For the fluctuations $\delta E_n^{(i,j)}$ with $j=5$ or $6$, to which we refer as ``light", one can also avoid explicit calculations using a simple trick illustrated in Fig.1 and Fig.2. Let us for example consider Fig.\ref{pic29}. The prescription of \cite{Gromov:2008bz} says that for the (2,9) and (1,10) fluctuations the residue
should be doubled compared to other excitations. By that reason we depict them by a double line.
This configuration can be thought of as adding a (2,5) pole
(and automatically a ``mirror'' (6,9) pole, due to (\ref{q-q})), plus a pole
between the second and the sixth sheets (and hence also a (5,9) pole). The poles on the
intermediate sheets have opposite residues and cancel. Both configurations on the right hand side of Fig.\ref{pic29} are equivalent and correspond
to the fermionic fluctuation with polarization (2,5). Hence we find that $\Omega_{25}(x)=\fr12\Omega_{29}(x)$.

At the next stage one has to solve the equation for the positions of the poles. From (\ref{q-q}) and (\ref{cpq-s})
it is clear that the solutions for the positions of the polarizations (2,5) and (2,9) are related by
\beq
x^{25}_{n}=x^{29}_{2n}\,.\label{double}
\eeq
Collecting all together we conclude that $\omega^{(2,5)}_n=\fr12\omega^{(2,9)}_{2n}$\,. The similar trick gives the remaining
frequencies $\omega^{(1,10)}_n$ and $\omega^{(3,7)}_n$ as one can see from the Fig.2.

\begin{figure}[t]
\label{pic37}
\epsfxsize=15cm
\centerline{\epsfbox{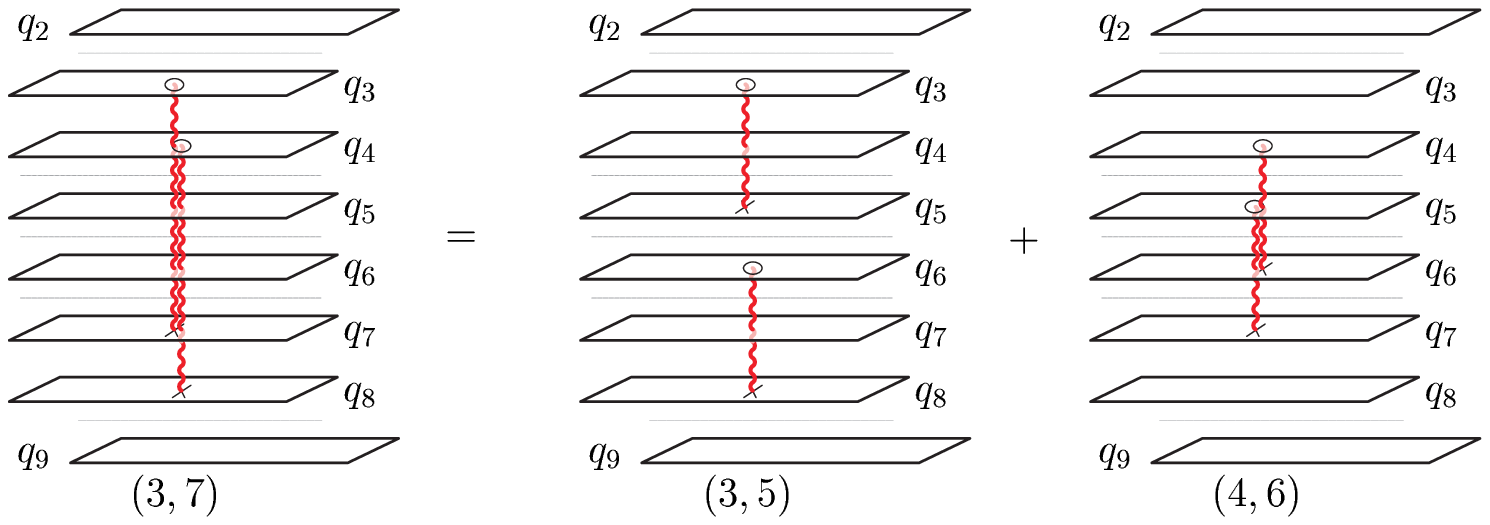}}
\caption{\small Equation relating fluctuations inside $\CP$ with polarizations (3,7) and (3,5). The (3,7) fluctuation can be decomposed into a (3,5) fluctuation plus
(4,7) fluctuation. The fluctuations (3,5) and (4,7) for the folded string are obviously equivalent.}
\end{figure}

\begin{table}[th]
\caption{\la{tab1}Notations for the frequencies of the folded string for $J\sim \log S\to\infty$.
The frequencies are taken from \cite{Frolov:2006qe}. In the table we use notations $\nu= \frac{J}{2\pi h},\;\frac{\nu}{\kappa}=\frac{\ell}{\sqrt{\ell^2+1}}$.
}
\beq \nn \bea{l|l}\toprule \rm\bf\quad\quad\quad eigenmodes& \rm \bf notation
\\ \midrule
    \bea{l}
    \sqrt{n^2+2\kappa^2\pm 2\sqrt{\kappa^4+n^2\nu^2}}\\
    \sqrt{n^2+2\kappa^2-\nu^2} \\
    \eea &
    \bea{l}
    \omega_n^{A_\pm}\\
    \omega_n^{A}\\
    \eea
\\ \midrule
     \sqrt{n^2+\kappa^2}  &
    \bea{l}
    \omega_n^{F}
    \eea
\\ \midrule
     \sqrt{n^2+\nu^2}&
    \bea{l}
    \omega_n^{S}
    \eea
\\ \bottomrule
\eea \eeq
\end{table}
\begin{table}[th]
\caption{\la{tab2}$3+8+5$ fluctuations around the folded string solution with multiplicities. Polarizations indicate which pair of sheets should be perturbed by a tiny pole.
}
\beq \nn
\bea{c|l|l|l}
\toprule
&{\rm \bf frequency} & {\rm \bf multiplicity} & {\rm \bf polarizations}\\
\midrule
{\rm \bf AdS}
& \bea{l}\omega^{A_+}_n\\ \omega^{A_-}_n\\ \omega^A_n \eea
& \bea{l}\times 1\\ \times 1\\ \times 1 \eea
& \bea{l}(1,10)\\ (2,9)\\ (1,9) \eea\\
\midrule
{\rm \bf fermions}
& \bea{l}\omega_n^F\\\omega_{2n}^{A_+}/2\\\omega_{2n}^{A_-}/2 \eea
& \bea{l}\times 4 \\\times 2\\\times 2 \eea
& \bea{l}(1,7);(1,8);(2,7);(2,8)\\ (1,5);(1,6)\\ (2,5);(2,6) \eea\\
\midrule
\CP
&\bea{l}\omega_n^S\\\omega_{2n}^{S}/2 \eea & \bea{l}\times 1 \\\times 4 \eea
&\bea{l} (3,7) \\ (3,5);(3,6);(4,5);(4,6) \eea
\\
\bottomrule
\eea
\eeq
\end{table}
\section{Summation Prescription}
Having the fluctuation frequencies $\omega_n^{(i,j)}$
computed we are ready to compute the shift of the
classical value of the energy due to  zero point oscillations.
Formally one can write
\beq
\delta E_{1-loop}=\frac{1}{2}\sum_{(i,j)} (-1)^{F_{(i,j)}}\sum_n \omega_n^{(i,j)}\;,
\eeq
however each sum over $n$ for a given polarization $(i,j)$ is divergent. A naive regularization which indeed leads to
a finite result is to interchange the order of summation over $n$ and $(i,j)$. Then the cancelations
between fermions and bosons make  the sum over $n$ convergent.

In the context of $\AD$ this regularization
leads to the agreement with the Bethe ansatz
prediction. However, repeating the same procedure
in $\Ad$ leads to a disagreement \cite{McLoughlin:2008ms,Alday:2008ut,Krishnan:2008zs}
with the prediction from the all-loop
Bethe ansatz \cite{Gromov:2008qe}.
In this section we will argue that this regularization is
not natural for the case of $\Ad$.
We present another regularization
which leads to a different finite result.
We argue by different means that our regularization is indeed
the one which should be used.

The main difference between $\Ad$ and $\AD$ theories
is existence of two dispersion relations in the BMN spectrum. Accordingly the BMN spectrum \cite{Nishioka:2008gz,Gaiotto:2008cg,Grignani:2008is} of $\Ad$ is naturally divided into two groups. Because of the differences in the diameters of $AdS_4$ and $\CP$ there are four fermions, three $AdS$ fluctuations and one $\CP$ fluctuation which
have the energies
\beq
{\cal E}_n=\frac{1}{\kappa}\sqrt{n^2+\kappa^2}\;,
\eeq
whereas the other four $\CP$ and four fermionic BMN  fluctuations have the following spectrum
\beq
\epsilon_n=\frac{1}{2\kappa}\sqrt{4n^2+\kappa^2}\;.
\eeq
In the Bethe ansatz language the first group of the fluctuations has both momentum carrying roots $u_4$ and $u_{\bar 4}$ excited, while the second group has only one of them. We call them heavy and light excitations respectively.
For even $n$ one can think about the fluctuations from the first group to be some kind of bound state of two ``light" excitations (with zero binding energy):
\beq
\E_{n}=\epsilon_{n/2}+\epsilon_{n/2}\;.
\eeq
Indeed, from this argument we see that for even $n$
``heavy" fluctuations with mode number $n$
and the ``light" fluctuations with mode number $n/2$
belong to the same family and should be treated together.
Looking at the fluctuations around the folded string
Tab.\ref{tab2} we also see that they are clearly separated into two groups\footnote{Formally one can pass to the BMN limit $\nu=\kappa$
to distinguish them.}. We define
\beq
K_n=\left\{
\bea{ll}
\omega_n^{\rm heavy}+\omega_{n/2}^{\rm light}&n\in {\rm even}\\
\omega_n^{\rm heavy}&n\in {\rm odd}
\eea
\right.
\eeq
where $\omega_n^{\rm heavy}$ and $\omega_n^{\rm light}$ in general are defined by
\beqa\la{hl}
\omega_n^{\rm heavy}&=&\omega_n^{(1,10)}+\omega_n^{(2,9)}+\omega_n^{(1,9)}
-\omega_n^{(1,7)}-\omega_n^{(1,8)}
-\omega_n^{(2,7)}-\omega_n^{(2,8)}
+\omega_n^{(3,7)}\\
\omega_n^{\rm light}&=&\omega_n^{(3,5)}+\omega_n^{(3,6)}+\omega_n^{(4,5)}
+\omega_n^{(4,6)}-\omega_n^{(1,5)}
-\omega_n^{(1,6)}-\omega_n^{(2,5)}
-\omega_n^{(2,6)}\nn\;.
\eeqa
We claim that the one-loop shift of the string energy is given by
\beq
E_{\rm 1-loop}=\lim_{N\to\infty}\sum_{n=-N}^{N} \frac{K_n}{2\kappa}\;.
\eeq
Using the notations given in Tab.\ref{tab2} one can pass to the fluctuations listed in Tab.1. For example $\omega_n^{(1,10)}=\omega_n^{A_+}$ and $\omega_n^{(3,5)}=\omega_{2n}^{S}/2$. From \eq{hl} we get
\beqa
\omega_n^{\rm heavy}&=&\omega_n^{A_+}+\omega_n^{A_-}+\omega_n^{A}-4\omega_n^{F}+\omega_n^{S}\\
\omega_{n/2}^{\rm light}&=&2\omega_n^{S}-\omega_n^{A_+}-\omega_n^{A_-}\nn\;.
\eeqa
For large $\kappa$ we can replace the sum with an integral
\beq
E_{\rm 1-loop}\simeq\lim_{N\to\infty}\int_{-N}^{N} \frac{2\omega_n^{\rm heavy}+\omega_n^{\rm light}}{4\kappa}dn
=\int_{0}^{\infty} \(2\omega_n^A+\omega_n^{A_+}+\omega_n^{A_-}+4\omega_n^S-8\omega_n^{F}\)\frac{dn}{2\kappa}\;.
\eeq
We notice that exactly this integral was computed in \cite{Frolov:2006qe}. We immediately write the result
\beq
E_{\rm 1-loop}\simeq \frac{J}{\ell}\frac{\sqrt{\ell^2+1}-1+2(\ell^2+1)\log\(1+\frac{1}{\ell^2}\)-(\ell^2+2)\log\frac{\sqrt{\ell^2+2}}
{\sqrt{\ell^2+1}-1}}{4\pi h\sqrt{\ell^2+1}}\la{res}
\eeq
where we used that $\frac{\nu}{\kappa}=\frac{\ell}{\sqrt{\ell^2+1}}$ and $\nu=\frac{J}{2\pi h}$. \eq{res} agrees completely with \eqs{fym}{f1loop}!
In particular, when $\ell\to 0$ we get
\beq
E_{\rm 1-loop}\simeq-\frac{3\log 2}{2\pi }\log S\;.
\eeq

\section{Summary}
In this paper we propose a particular regularization of the sum over
the fluctuation frequencies. The regularization goes as follows: we split
the fluctuations into two groups with $4$ bosons and $4$ fermions each.
We call them heavy and light fluctuations since for even mode numbers $n$ heavy ones
can be decomposed  into a sum of two light fluctuations with mode number $n/2$. The light fluctuations are in some sense more fundamental. We prescribe then the mode number $2n$ heavy fluctuations to be treated together with mode number $n$ light fluctuations.
Following this prescription we match the one-loop energy shift with
the prediction from the all-loop Bethe equations.
This regularization method looks a bit artificial on the $n$ plane,
but from the algebraic curve point of view it makes perfect sense.
The positions of the excitations on the algebraic curve of
the light and heavy excitations are related in the following way
\beq
x^{\rm heavy}_{2n}= x_n^{\rm light}
\eeq
(see for example \eq{double}). We can regularize the sum over the fluctuations in
terms of the algebraic curve by introducing an $\epsilon$ balls around $x=\pm 1$ and
taking into account only the fluctuations living in their exterior.
For the $\AD$ string this $\epsilon$ regularization leads to the simple
cut-off prescription in the sum over fluctuations. However, in the $\Ad$ case one should be more careful.

Whereas it is clear from the algebraic curve why this particular regularization
makes sense it is rather hard to justify it starting from the world-sheet string action.
At the end of the day the algebraic curve and the action are two equivalent descriptions
of the semiclassical strings and it should be possible to understand this regularization from both points of view.
One of the possible approaches to better understanding of the problem is to pass to the
Frolov-Tseytlin limit ($J\to\infty$ and $J/h$ large) where the conjectured all-loop Bethe equations reduce to the two-loop
Bethe equations derived from the $CS$ perturbation theory \cite{Minahan:2008hf}. Our hope is that the finite-size
corrections to the scaling limit of the two-loop Bethe equations will only be consistent with the sum over fluctuations if our prescription is used.

\section*{Acknowledgments}
We would like to thank L.~Alday, G.~Arutyunov, C.~Krishnan, I.~Kostov, T.~McLoughlin, R.~Roiban, S.~Schafer-Nameki, D.~Serban, I.~Shenderovich, A.~Tirziu, A.~Tseytlin, D.~Volin, K.~Zarembo and especially P.~Vieira for many useful discussions. V.M. would also like to thank A.~Gorsky for introducing into the topic of integrability in AdS/CFT. NG was partially supported by RSGSS-1124.2003.2, by RFBR grant
08-02-00287 and ANR grant INT-AdS/CFT (contract ANR36ADSCSTZ). The work of V.M. was partially supported by RFFI grant 06-02-17382, grant for support of scientific schools NSh-3036.2008.2 and CNRS-RFBR grant PICS-07-0292165. This work was done during our stay at Les Houches Summer School, which we thank for hospitality.

\newpage
\bibliographystyle{utphys}
\renewcommand{\refname}{Bibliography}
\addcontentsline{toc}{section}{Bibliography}
\bibliography{NOTEbib}


\end{document}